\documentclass[final,5p,times,twocolumn,authoryear]{elsarticle}
\graphicspath{ {./figures/} }
\usepackage{csquotes}
\usepackage{float}
\usepackage{listings}

\usepackage[T1]{fontenc}
\usepackage{verbatim} 
\usepackage{apalike}
\usepackage{graphicx}
\usepackage[figuresright]{rotating}
\usepackage{subcaption}
\usepackage{array}
\usepackage{caption}
\usepackage{tabularx}
\usepackage{amsmath}
\usepackage{float}
\usepackage{booktabs}
\usepackage{geometry}
\usepackage{pdflscape}
\usepackage{afterpage}
\usepackage{rotating}
\restylefloat{figure}
\restylefloat{table}
\usepackage{hyperref}
\usepackage{xcolor} 
\hypersetup{
    colorlinks=true,
    linkcolor=blue,
    citecolor=blue,
    urlcolor=blue,
}
\usepackage{amssymb}
\usepackage{lipsum}

\journal{Expert systems with applications}

\begin{document}

\begin{frontmatter}

\title{Jup2Kub: algorithms and a system to translate a Jupyter Notebook pipeline to a fault-tolerant distributed Kubernetes deployment}

\author[first]{Jinli Duan}
\affiliation[first]{organization={New York University, Tandon School of Engineering},
            city={New York},
            state={NY},
            country={United States}}
            \ead{jd5374@nyu.edu}
\cortext[cor1]{Corresponding author.}
\author[second]{Dennis Shasha\corref{cor1}}
\affiliation[second]{organization={New York University},
            city={New York},
            state={NY},
            country={United States}}
            \ead{shasha@cims.nyu.edu}

\begin{abstract}
Scientific workflows  orchestrate a series of computational, data manipulation, and sometimes visualization steps to analyze scientific data. They encapsulate the essential processes and relevant data required to reproduce and validate experiments, typically comprising computational steps for scientific simulations or data analysis. 
Domain scientists often develop such workflows as Jupyter notebooks. Such notebooks are convenient and powerful, but often encounter three problems: 1. they don't scale with increasing sizes of data; 2. they are not tolerant to failures; 3. they stop working when underlying tools and packages change. Jup2Kup is a software system designed to translate scientific workflows from Jupyter notebooks to a distributed, high-performance environment on Kubernetes, thereby encapsulating fault-tolerance features. The envisioned system also captures software dependencies to ensure that the system continues to work even as underlying tools change. 
\end{abstract}


\end{frontmatter}

\section{Introduction}

Generally speaking, reproducibility is one of the fundamental principle in research, yet remains challenging to achieve in practice. Three main issues get in the way: source code is often not shared openly, details on computing systems are lacking, and experimental procedures are not reported comprehensively\cite{jup4workflow}. When researchers incorrectly sharing the code, specify their software environments setting, and document their parameters, simulation and analysis methods, it becomes almost impossible to replicate their results. This highlights the need for more reproduced, transparent research software framework for scientific workflow.

The Jupyter Notebook has revolutionized scientific workflows in research field. Its interactive web interface combines code, text, visualizations, and other media into single sharable documents. This integration enhances efficiency, reproducibility, and knowledge sharing across research domains. Jupyter's support for multiple programming languages through interchangeable kernels makes it versatile for diverse projects. Jupyter uses the "one study - one document" structure\cite{jup4workflow} ensures a comprehensive, executable record of the research process, improving reproducibility and transparency. This makes Jupyter Notebook uniquely useful for managing computational workflows from initial exploration to final publication. Its shareable, interactive format supports diverse applications from documentation to domain-specific languages. 

While Jupyter Notebooks offer many advantages in scientific workflows, they also have limitations. One significant limitation is the lack of modularity. Jupyter tends to encourage writing code directly into cells, which can hinder the development of modular code structures crucial for large-scale projects. This structure makes it challenging to transition larger projects into more reusable and maintainable code modules.\cite{datascientistsnotebook} The linear nature of notebooks can lead to challenges in managing complex computations, as dependencies and data are tightly mixed across the notebook's flow. Moreover, many scientists using Jupyter are experts in their fields but not necessarily in computer science\cite{bettercode}. This issue often results in notebooks with less-than-optimal code quality and practices\cite{bettercode}. 

Given the limitations of Jupyter Notebooks, particularly in terms of their linear workflow and the varied expertise of users, there's a clear need for a more advanced software development framework. Such a framework should not only offer higher fault tolerance and robust error-handling mechanisms but also enhance modularity. This combination would accommodate users with different levels of programming skills more effectively. Improved modularity would lead to better code organization and easier maintenance, while enhanced fault tolerance would ensure more reliable and accurate outcomes in computational tasks.

Jup2Kub represents an innovative strategy for integrating Jupyter Notebooks into Kubernetes clusters, proposing a method to divide notebooks into distinct processes and adapt them for deployment in Kubernetes pods. It does this by breaking down notebooks into individual processes, wrapping them in Kubernete pods—self-contained environments that have all the necessary pieces to run the code. The system ensures that even if one part fails, the workflow can continue, thanks to built-in fault tolerance features like automatic retries and backups.This concept underscores the importance of containerization for each process and highlights the system's enhanced fault tolerance capabilities.

Through this effort, we aim to provide researchers with a tool that simplifies the complex aspects of distributed computing, letting them focus entirely on their research work. With Jup2Kub, our goal is to give the scientific community a user-friendly tool that removes the complexities of distributed computing, allowing them to concentrate on their core research. This initiative is designed to improve the efficiency and reliability of scientific workflows, ensuring that data analysis and simulations can be conducted effectively and reliably on a large scale.

\section{Related Work}
Before diving into the details of our software design, it's useful to review related prior work that influenced our approach. Jupyter notebooks have become prominent for scientific computing due to integrating code, results, and narrative. However, making notebooks' workflows parallel, fault-tolerant, and robust remains challenging. Many projects have explored solutions like containerizing notebooks or integrating cluster managers to enable large-scale reproducible research. We build on these efforts with an innovative Kubernetes-based approach to distribute notebooks while retaining interactivity. Our goal is to advance the state of the art by combining the best of previous solutions. This context of related work provides background on how we aimed to move notebook-based scientific computing forward with our framework.

\subsection{Reprozip}
ReproZip is an open-source tool designed to enhance the reproducibility of a wide range of applications, from data analysis tools and scripts to client-server applications and Jupyter notebooks.  It operates by tracing system calls to automatically identify necessary files for an experiment, allowing users to review and modify this list and metadata before creating a package. These packages can then be reproduced using various methods such as chroot environments, Vagrant-built virtual machines, and Docker containers, with the flexibility to add more options through plugins.\cite{reprozip}

In Jup2Kub, ReproZip plays an important role, particularly in the step of dockerlizing Jupyter Notebook. By tracing system calls used during the notebook, ReproZip  identifies the necessary files to be included in the Docker container. This ensures not only the completeness of required files but also preserves the configuration and dependencies of the runtime environment. This step significantly enhances the portability and reproducibility of our software and is a critical component of our design strategy.
\subsection{Kubernetes}
Kubernetes, also known as K8s, is an open-source platform designed for automating the deployment, scaling, and management of containerized applications. Through Kubernetes, complex workflows can be containerized and managed efficiently across diverse computing environments. This containerization process involves encapsulating each component of a workflow within containers, leading to improved portability, scalability, and resource utilization. \cite{kuberneteswork}

Kubernetes pods is the most important concept for this system. Pods are a fundamental element in Kubernetes and represent the smallest deploy-able units that can be created, scheduled, and managed.\cite{kubernetessurvey}

Built on Kubernetes, KubeAdaptor is an innovative cloud-native scientific workflow framework optimized through the scheduling algorithm. It streamlines integrating workflow systems with Kubernetes for consistent task scheduling.By efficiently coordinating containerized workflows on Kubernetes, KubeAdaptor has it's own approach of advances Kubernetes-based workflow schedule.\cite{kubeadaptor}
\subsection{Pegasus}
Pegasus is a widely adopted scientific workflow manager valued for automating complex computational workflows. It enables researchers to describe workflows abstractly, which Pegasus maps to distributed resources efficiently. This capability to map workflows onto distributed environments allows Pegasus to streamline executing large-scale tasks by handling data transfers and optimizations under the hood.

In designing Jup2Kub, we drew inspiration from Pegasus's structure and capabilities. Specifically, Pegasus demonstrated the potential of leveraging distributed systems to efficiently manage workflow execution without the user needing to handle low-level details. Following this model, Jup2Kub taps into the power of Kubernetes to automate the execution of Jupyter notebook workflows across clusters. Kubernetes serves as the distributed platform for deploying notebook containers and workloads in a scalable fashion.

By abstracting these distributed computing complexities, Jup2Kub, like Pegasus, allows researchers to focus on workflow logic rather than infrastructure\cite{pegasus}. Building on the strengths of Pegasus for workflow mapping and execution, Jup2Kub aims to bring similar automation and distribution capabilities natively to the popular Jupyter notebook environment for scientific computing.

\begin{figure*}
  \centering
  \includegraphics[width=0.75\textwidth]{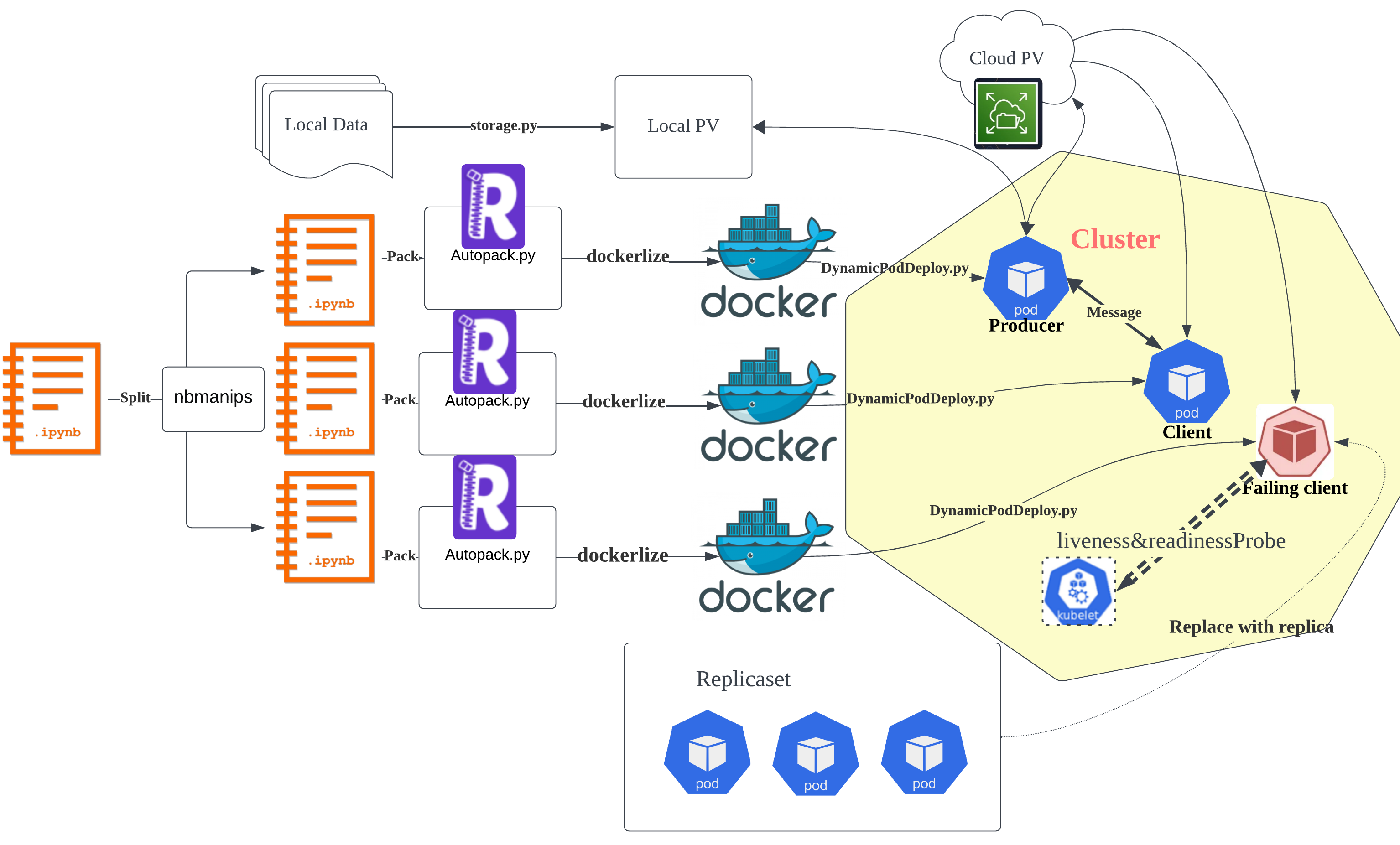}
  \caption{Flowchart}
  \label{fig:figure1}
\end{figure*}

\section{Software Architecture}
Jup2Kub aims to automate the segmentation of Jupyter notebooks into distinct executable steps, package these steps into Docker containers, and orchestrate the deployment and execution on a Kubernetes cluster as Kubernete pods. By utilizing Kafka for inter-pod communication and implementing ReplicaSets, liveness and readiness probes, and other Kubernetes-native features, the system is engineered to offer fault tolerance. Additionally, shared storage solutions using Persistent Volumes (PV) and Persistent Volume Claims (PVC) are integrated to ensure data consistency and availability across the distributed setup.
\subsection{Split and Dockerlize}
In this section, Jup2Kub starting by using ReproZip's packaging feature to capture the necessary dependencies for the experiment as config.yml. Then, using the nbmanips library, Jup2Kub splits the Jupyter Notebook based on its piped sections. Following this, each split Notebook is individually packaged by the Reprozip's pack command, substituting the dependencies of each with those initially captured. Finally, Jup2Kub unpacks these packaged Notebooks into individual Docker containers. This process achieves effective segmentation and containerization of the Notebook, enhancing workflow management and execution.
\subsection{Dynamic Pod Deloyment}
This section outlines a Python framework for systematically deploying pods in Kubernetes. A PodManager class handles pod details like producer/consumer type and topics. A DynamicPodDeployer class then deploys pods by generating and applying Kubernetes configurations defining environment variables, persistent volumes (PVs), persistent volume claims (PVCs), and strategies for fault tolerance and optimization. The script incorporates error handling and logging for smooth deployment and troubleshooting. This demonstrates an efficient, scalable approach to pod deployment and management in Kubernetes, useful for cloud-based applications.
\subsubsection{PodManager Class}
The PodManager class handles pod details like producer/consumer type and associated topics for each pod.
\subsubsection{Loading K8s config}
DynamicPodDeployer initializes by loading the Kubernetes configuration using config.load-kube-config().
\subsubsection{DynamicPodDeployer Class}
The deployer iterates through all pods from PodManager, deploying each to Kubernetes. It tailors the deployment per pod role (producer/consumer) and configures environment variables accordingly. We will be using the following listing as a pod depolyment template:
\begin{lstlisting}[language=Python, caption={Kubernetes pod Deployment template}, label={lst:k8s-deployment}]
apiVersion: apps/v1
kind: Deployment
metadata:
  name: {pod_name}-deployment
spec:
  replicas: 3
  strategy:
    type: RollingUpdate
    rollingUpdate:
      maxUnavailable: 1
      maxSurge: 1
  selector:
    matchLabels:
      app: {pod_name}
  template:
    metadata:
      labels:
        app: {pod_name}
    spec:
      containers:
      - name: {pod_name}-container
        image: {image_name}:{tag}
        env:
        - name: KAFKA_BROKER
          value: "my-broker-address"
        {env_content}
        resources:
          limits:
            cpu: "1"
            memory: "1Gi"
          requests:
            cpu: "500m"
            memory: "500Mi"
        livenessProbe:
          httpGet:
            path: /healthz
            port: 8080
        readinessProbe:
          httpGet:
            path: /readiness
            port: 8080
        volumeMounts:
        - name: efs-volume
          mountPath: /mnt/efs
      volumes:
      - name: efs-volume
        persistentVolumeClaim:
          claimName: {pod_name}-efs-pvc
\end{lstlisting}
\subsection{Storage}
Jup2Kub leverages Kubernetes Persistent Volumes (PVs) and Persistent Volume Claims (PVCs) for storage handling. For local storage, it creates PVs and PVCs tied to specific nodes, useful for workloads requiring fast node-level access. For cloud storage, it utilizes AWS Elastic Block Store (EBS) and Elastic File System (EFS) to create networked PVs and PVCs that enable shared, scalable storage across nodes.

Locally, PVs are configured with a storage path and capacity on each node. PVCs request these resources from a namespace. Node affinity binds PVs to nodes, allowing low-latency access to local files.

For the cloud, PVs and PVCs are defined with configurations for EBS and EFS. This facilitates elastic storage and data sharing in the cloud.

Combined, these local and cloud-based PVs and PVCs deliver efficient, reliable storage management across infrastructure types for Jup2Kub's data-intensive workflows. The multi-tiered storage architecture adapts to workload needs.
\subsection{interprocess communication}

This "Communication" section describes a process for interacting with services within a Kubernetes cluster. The script defines two main functions:

get-cluster-ip(): This function retrieves the cluster IP of a specified service within a given namespace using kubectl. It uses subprocess to execute the command and capture the output, which is the cluster IP.

communicate-with-service(): This function uses the obtained cluster IP to communicate with the service. It constructs a URL and makes an HTTP GET request to the service using the requests library. If successful, it returns the response text; otherwise, it handles potential exceptions and returns None.

The main script takes user input for the service name and namespace, retrieves the service's cluster IP, and then attempts to communicate with the service using this IP\cite{k8scommuni}. It prints the response from the service or an error message if communication fails.
\subsection{Fault Tolerance Management}
Fault tolerance mechanisms are implemented in Kubernetes and has been adopted for this system\cite{k8sarch}. Replicasets, liveness probes, readiness probes, rolling Updates, horizontal Pod autoscalers, comprehensive error handling and logging, together, these strategies provide a robust, self-healing infrastructure to sustain workflow execution despite disruptions. The multi-layered approach delivers resilience through redundancy, automatic scaling, health checks, and observability.

\begin{itemize}
  \item \textbf{ReplicaSet:} Sets the number of replicas to 3, ensuring three nginx pods are maintained for backup\cite{k8sarch}.
  \item \textbf{Resource Limits:} Defines CPU and memory limits for each replica to optimize resource usage.
  \item \textbf{Liveness and Readiness Probes:} These probes monitor the health of pods, replacing them with replicas if they become unresponsive.
  \item \textbf{Rolling Updates:} Controlled via `maxSurge` and `maxUnavailable`, these updates allow for continuous service during application updates\cite{Rollingupdate}.
  \item \textbf{Horizontal Pod Autoscaler:} Dynamically adjusts CPU and GPU resources based on workload demands.
  \item \textbf{Error Handling and Logging:} Robust mechanisms for error tracking and logging, essential for system monitoring and troubleshooting.
\end{itemize}

\section{Conclusion}
This report introduced Jup2Kub, a framework design to simplify containerizing and deploying Jupyter Notebooks on Kubernetes. Jup2Kub and its tools streamline complex workflow management by enabling easy, efficient handling of containerized tasks. By lowering barriers to deployment and scaling, Jup2Kub promotes more accessible, reproducible research. With such capabilities, we hope to empower the scientific community to share code and data openly, propelling reproducible findings and collaborative progress. Jup2Kub paves the way for workflows that are robust yet flexible, furthering the impact of computational research.

\bibliographystyle{elsarticle-harv} 
\bibliography{Jup2Kub}

\end{document}